\documentclass[amsmath,amssymb,prl,hyperlink,twocolumn]{revtex4}

\usepackage{graphicx}
\usepackage{soul}
\usepackage[colorlinks=true,citecolor=blue,linkcolor=magenta]{hyperref}
\usepackage[usenames]{color}
\usepackage{amsfonts}
\usepackage{color}

\def\ket#1{\left\lvert {#1} \right\rangle}

\begin{document}

\title{Experimental quantum computing to solve systems of linear equations}

\author{X.-D. Cai$^1$, C. Weedbrook$^2$, Z.-E. Su$^{1}$, M.-C. Chen$^1$, Mile Gu$^{3,4}$, M.-J. Zhu$^1$, Li Li$^{1}$, Nai-Le Liu$^{1}$, \\ Chao-Yang Lu$^{1}$, Jian-Wei Pan$^1$\vspace{0.2cm}}

\affiliation{$^1$ Hefei National Laboratory for Physical Sciences at Microscale and Department of Modern Physics, University of Science and Technology of China, Hefei, Anhui 230026, China}
\affiliation{$^2$ Center for Quantum Information and Quantum Control,
Department of Electrical and Computer Engineering and Department of Physics, University of Toronto, Toronto, M5S 3G4, Canada}
\affiliation{$^3$ Centre for Quantum Technologies, National University of Singapore, Singapore}
\affiliation{$^4$ Center for Quantum Information, Institute for Interdisciplinary Information Sciences, Tsinghua University, Beijing, China}

\date{\today}

\begin{abstract}
Solving linear systems of equations is ubiquitous in all areas of science and engineering. With rapidly growing data sets, such a task can be intractable for classical computers, as the best known classical algorithms require a time proportional to the number of variables $N$. A recently proposed quantum algorithm shows that quantum computers could solve linear systems in a time scale of order $\log(N)$, giving an exponential speedup over classical computers. Here we realize the simplest instance of this algorithm, solving $2\times2$ linear equations for various input vectors on a quantum computer. We use four quantum bits and four controlled logic gates to implement every subroutine required, demonstrating the working principle of this algorithm.
\end{abstract}

\maketitle

The problem of solving a system of linear equations plays a central role in diverse fields such as signal processing, economics, computer science, and physics. Such systems often involve tera or even petabytes of data, and thus the number of variables $N$, is exceedingly large. However, the best known algorithms for solving a system of $N$ linear equations on classical computers requires a time complexity on the order of $N$, posing a formidable challenge. 

Harnessing the superposition principle of quantum mechanics, quantum computers \cite{Nielsen2000, ObrienNatureReview} promise to provide exponential speedup over their classical counterparts for certain tasks. Notable examples include quantum simulation \cite{Feynman1982,simulationLloyd} and Shor's quantum factoring algorithm \cite{Shor1997}, which have driven the field of quantum information over the past two decades as well as generating significant interest in quantum technologies that have enabled experimental demonstrations of the quantum algorithms in different physical systems \cite{shorNMR, shorPhoton, photonSimulation, trappedionSimulation, shorSuperconducting}.

Recently, Harrow, Hassidim and Lloyd~\cite{Harrow2009} proposed another powerful application of quantum computing for the very practical problem of solving systems of linear equations. They showed that a quantum computer can solve a system of linear equations exponentially faster than a classical computer in situations that we are only interested in expectation values of an operator associated with the solution rather than the full solution. A quantum algorithm has been designed such that the value of this property may be estimated to any fixed desired accuracy within $O(\log(N))$ time, making it one of the most promising applications of quantum computers.

In this article, we report an experimental demonstration of the simplest meaningful instance of this algorithm, that is, solving $2\times2$ linear equations for various input vectors. The quantum circuit is optimized and compiled into a linear optical network with four photonic quantum bits (qubits) and four controlled logic gates, which is used to coherently implement every subroutine for this algorithm. For various input vectors, the quantum computer gives solutions for the linear equations with reasonably high precision, ranging from fidelities of $0.825$ to $0.993$.

\begin{figure*}[tb]
    \centering
        \includegraphics[width=0.68\textwidth]{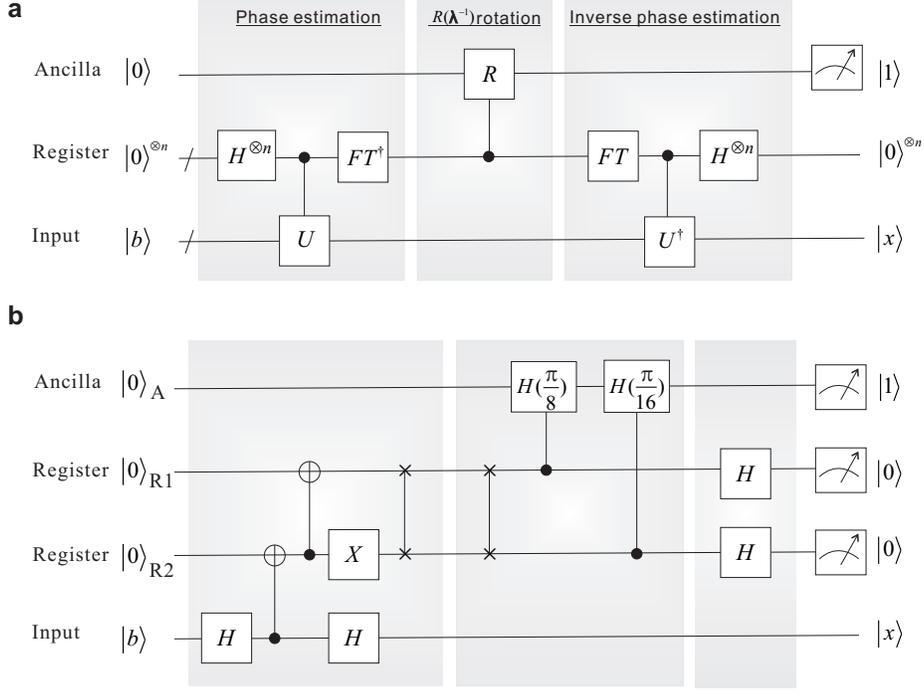}
\caption{Quantum circuits for solving systems of linear equations. \textbf{a}, Outline of the original quantum algorithm proposed in~\cite{Harrow2009}. The light gray blocks represent three basic subroutines of the algorithm. $U=\sum_{k=0}^{T-1}{|k\rangle}{\langle k|}\otimes e^{iAkt_0/T}$, where $T=2^t$ with $t$ being the number of registers, and $t_0$ is chose as $2\pi$. H is a Hadamard gate.
$FT$ and $FT^{\dagger}$ are the Fourier transformation and the inverse Fourier transformation \cite{Nielsen2000}, respectively. The controlled rotation $R$ evolves the system into the state $\sum_{j=1}^N \beta_{j}|u_{j}\rangle|\lambda_{j}\rangle (\surd{(1-C^2/\lambda_{j}^2)|0\rangle+({C}/{\lambda_{j}})|1\rangle)}$, where C is a normalizing constant. The inverse phase estimation subroutine restores the register to $|0\rangle^{\otimes n}$. Finally, the vector $\vec{x}$, the solution of the system of linear equations, can be obtained by conditioning on the measurement outcome of $|1\rangle$ in ancilla qubit. \textbf{b}, The optimized circuit with four qubits and four entangling gates (see main text for details). The two swap gates are canceled out.}
\label{fig1}
\end{figure*}
%
%


The problem of solving linear equations can be summarized as follows: We aim to solve $A \vec{x} = \vec{b}$ for $\vec{x}$, when given a $N \times N$ Hermitian matrix $A$ and a vector $\vec{b}$. To adapt this problem to quantum processing, $\vec{x}$ and $\vec{b}$ are scaled to unit length (i.e., $||\vec{x}|| = ||\vec{b}|| = 1$). Thus, a vector $\vec{b}$ can be represented by a quantum state $\ket{b} = \sum_i b_i \ket{i}$ on $O(\log (N))$ qubits where $\ket{i}$ denotes computational basis. The desired solution $\vec{x}$ can then be encoded within the quantum state as
\begin{equation}
\ket{x} =  c A^{-1} \ket{b} ,\qquad c^{-1} = ||A^{-1} \ket{b}||.
\end{equation}
The quantum algorithm devised in ref.~\cite{Harrow2009} was designed to synthesize $\ket{x}$ (see Fig.~\ref{fig1}a). The quantum algorithm involves three subsystems: a single ancilla qubit initialized in $\ket{0}$, a register of $n$ qubits of working memory initialized in $|0\rangle^{\otimes n}$ and an input state initialized in $\ket{b}$. The input state $\ket{b}$ can be expanded in the basis of $\ket{u_j}$ as $\ket{b} = \sum_{j=1}^N \beta_j \ket{u_j}$, where $\ket{u_j}$ is eigenstate of $A$, and $\beta_j = \langle {u_j}|b\rangle$. Execution of the algorithm can be decomposed into three subroutines: (1) phase estimation, (2) controlled rotation and (3) inverse phase estimation.

Step (1) is used to determine the eigenvalues of $A$, which we denote by $\lambda_j$. Phase estimation is essentially a controlled unitary with a change of basis that maps the eigenvalues onto the working memory \cite{Nielsen2000,phaseEstimation}. The phase estimation protocol is applied to the input, using the working memory as control, to give
\begin{align}
\sum_{j=1}^{N}\beta_{j}|u_{j}\rangle|{\lambda}_{j}\rangle,
\end{align}
where $\ket{\lambda_j}$ represents the binary representation of $\lambda_j$, stored to a precision of $n$ bits.

In step (2), one needs to extract the eigenvalues of  $A^{-1}$, i.e. $\lambda_j^{-1}$ from $\ket{\lambda_j}$. This is realized through an additional ancillary qubit initialized in the state $\ket{0}$. Application of an appropriate controlled rotation $R({\lambda}^{-1})$ on this qubit (see Fig.~\ref{fig1}a) transforms the system to
\begin{align}
\sum_{j=1}^N \beta_{j}|u_{j}\rangle|\lambda_{j}\rangle \big(\sqrt{1-\frac{C^2}{{\lambda_{j}}^2}}|0\rangle
+\frac{C}{\lambda_{j}}|1\rangle \big).
\end{align}
The final step involves applying the gate sequence of step (1) in reverse. This disentangles the register, which is reset to  $|0\rangle^{\otimes n}$. Therefore we end up with
\begin{align}
\sum_{j=1}^N \beta_{j}|u_{j}\rangle \big(\sqrt{1-\frac{C^2}{{\lambda_{j}}^2}}|0\rangle
+\frac{C}{\lambda_{j}}|1\rangle\big).
\end{align}
Measurement of the ancillary qubit and post-selection (with a successful probability) of an outcome of $|1\rangle$ will result in an output state $\sum_{j=1}^N C({\beta_{j}}/{\lambda_{j}})|u_{j}\rangle$ which is proportional to our expected result state $\ket{x}$.

Resource needed for the algorithm of a general $s$-sparse $N \times N$ matrix $A$ is estimated to be $O(log(N)s^2\kappa/\varepsilon)$, where $\kappa$ is the condition number (the
ratio between $A$'s largest and smallest eigenvalues), $\varepsilon$ is the acceptable error of the output vector (see \cite{Harrow2009} for more details). Putting together the success probability in the post-selection measurement in step (3), the total runtime of the quantum algorithm is $O(log(N)s^2{\kappa}^2/\varepsilon)$, which outperforms the best classical one and reach an exponential speedup generically  \cite{Harrow2009}.

\begin{figure*}[tb]
    \centering
        \includegraphics[width=0.68\textwidth]{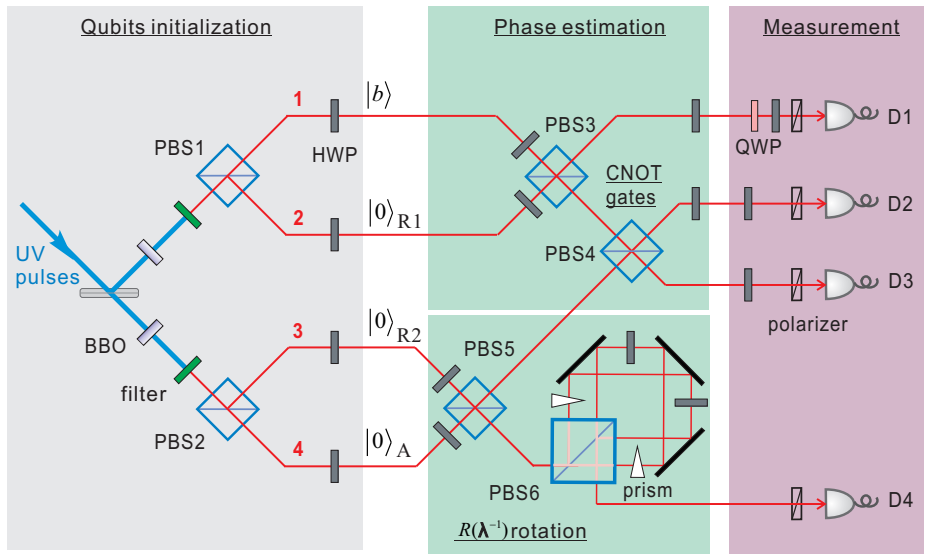}
\caption{Experimental setup. There are four key modules in the optical setup. (1) Qubit initialization: Ultraviolet laser pulses with a central wavelength of $394$~nm, pulse duration of $120$~fs and a repetition rate of $76$ MHz
pass through two $\beta$-barium borate (BBO) crystals to produce two photon pairs. The four single photons are spatially separated by PBS$_1$ and PBS$_2$ and initialized using HWPs, with three of them be in the state $|H\rangle_i$ where $i$ denoting
their spatial modes and one  be in state $|b\rangle$. Photon $1$ is used as the input vector qubit and photon $4$ is used as the ancilla.  Photons $2$ and $3$ are used as the register qubits $R1$ and $R2$, respectively.
(2). Phase estimation: The input qubit $|b\rangle$ is mixed with the two register qubits on PBS$_3$ and PBS$_4$ to simulate the CNOT gates in Fig.~\ref{fig1}b. (3) $R(\lambda^{-1})$ rotation:
An entanglement-based implementation of the two controlled unitary gates (see main text for details). (4) Inverse phase estimation: This is realized semiclassically by using measurement and classical feed forward.
To achieve good spatial and temporal overlap, all photons are spectrally filtered ($\lambda_{\textrm{FWHW}}=3.2$~nm) and detected by fiber-coupled single-photon detectors (D1, $\cdots$, D4). The coincidence events are registered by a programmable multichannel coincidence unit.}
\label{fig2}
\end{figure*}
\begin{figure*}[tb]
\centering\includegraphics[width=0.74\textwidth]{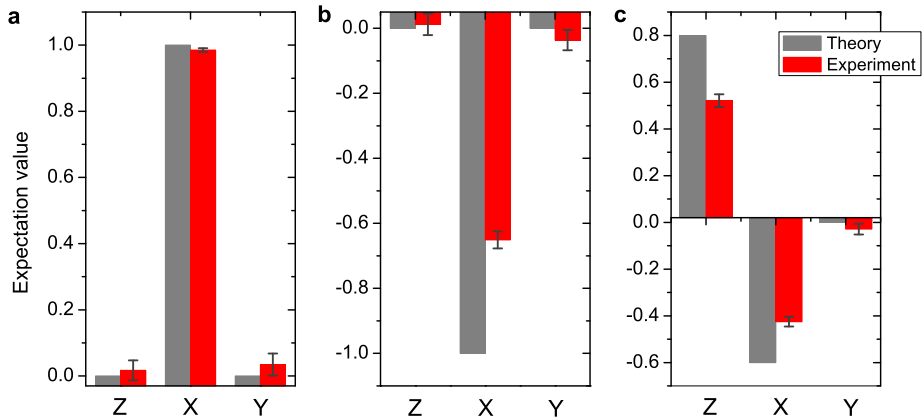}
\caption{Experimental results. Three different input vectors are chosen: \textbf{a}, $|b_{1}\rangle=(|H\rangle+|V\rangle)/\sqrt{2}$, \textbf{b}, $|b_{2}\rangle=(|H\rangle-|V\rangle)/\sqrt{2}$, and \textbf{c}, $|b_{3}\rangle=|H\rangle$. The quantum algorithm is run to determine the the expectation value $\langle x|\hat{M}|x\rangle$, where $\hat{M}$ is some operator. For each input state $|b\rangle$, the theoretically predicted (gray bar) and experimentally measured (red bar) expectation values of the observables of the Pauli matrices $Z$, $X$, and $Y$ are presented. The output states are measured with a fidelity of 0.993(3), 0.825(13), and 0.836(16) for $|b_{1}\rangle$, $|b_{2}\rangle$, and $|b_{3}\rangle$, respectively. The error bars denote one standard deviation, deduced from propagated Poissonian counting statistics of the raw detection events.}\label{fig3}
\end{figure*}
%
%


Here we demonstrate a proof-of-principle experiment of this algorithm: solving systems of $2 \times 2$ linear equations. We choose the matrix $A$ to be
\begin{align} A =
\left(\begin{array}{cc}
  1.5 & 0.5 \\
  0.5 & 1.5
\end{array}\right),
\end{align}
and we choose the following values for input vector $|b\rangle$
\begin{align}\ket{b_1} =  \frac{1}{\sqrt2}
 \left(
              \begin{array}{c}
               1 \\
                1 \\
              \end{array}
            \right),
\ket{b_2} =  \frac{1}{\sqrt2}
 \left(
              \begin{array}{c}
               1 \\
               -1 \\
              \end{array}
            \right),
\ket{b_3} =
 \left(
              \begin{array}{c}
               1 \\
                0 \\
              \end{array}
            \right).
\end{align}
The matrix $A$ is chosen such as its eigenvalues are $1$ and $2$ which can be encoded with two qubits in registers \cite{footnote}.
This allows us to optimize the circuit requiring four qubits and four entangling gates as shown in Fig.~\ref{fig1}b. The phase estimation subroutine of the circuit can be compiled into two controlled-NOT (CNOT) gates, a swap gate, and three single qubit rotation gates. Following the circuit design of Ref.~\cite{Cao2012}, the R(${\lambda}^{-1}$) rotation subroutine is implemented in two steps: finding the reciprocal $|1/\lambda_j\rangle$ from eigenvalue $|\lambda_j\rangle$ stored in registers, which in our case can be realized by a swap gate, and controlled unitary gates $H(\theta)$, where
\begin{equation}
\label{Hgate}
H(\theta)=
\left(\begin{array}{cc}
  \cos(2\theta) & \sin(2\theta) \\
  \sin(2\theta) & -\cos(2\theta)
\end{array}\right).
\end{equation}
Finally, the subroutine of the inverse phase estimation is realized using a semiclassical version that employs single-qubit rotations conditioned on measurement outcomes \cite{semiclassical}.

To implement the quantum circuit shown in Fig.~\ref{fig1}b, we prepare four single photons from spontaneous parametric down-conversion \cite{SPDC} as the input qubits (Fig.~\ref{fig2}). The horizontal (H) and vertical (V) polarizations of the single photons are used to encode the logic qubits $|0\rangle$ and $|1\rangle$, respectively. The experimental challenge of implementing the circuit in Fig.~\ref{fig1}b lies in the four entangling gates between the single photonic qubits.

In the phase estimation subroutine, noting that the target qubits of the CNOT gates are fixed, their implementations can be simplified using combinations of a polarization beam splitter (PBS) and a half-wave plate (HWP), through which an arbitrary control qubit $\alpha|H\rangle+\beta|V\rangle$ and the target qubit $|H\rangle$ evolve into $\alpha|H\rangle|H\rangle+\beta|V\rangle|V\rangle$ which is the desired output of CNOT operations \cite{photonCNOT1}. The R(${\lambda}^{-1}$) rotation subroutine involves two consecutive controlled unitary gates, H($\pi$/8) and H($\pi$/16). Instead of decomposing it into multiple CNOT gates \cite{Nielsen2000,photonCNOT1}, we adopt a more efficient, entanglement-based construction method \cite{Zhou}. The ancillla qubit is first entangled with the register qubits by mixing on PBS$_5$, and then passed through a polarization-dependent Sagnac-like interferometer where the desired controlled unitary operations are applied (see \cite{supplemental} for more details and photon loss analysis). Finally, the ancillary qubit is measured, and when an outcome state $|1\rangle$ is obtained, the algorithm is announced successful.


Before running the algorithm, we first characterized the performance of the optical quantum circuit. The two registers, ancilla and input qubits ($|b_3\rangle$) are initialized in the $|H\rangle_A\otimes|H\rangle_{R1}\otimes|H\rangle_{R2}\otimes|H\rangle_b$ state. Theoretically these four qubits will evolve into a maximally-entangled four-qubit Greenberger-Horne-Zeilinger~(GHZ) state during the subroutine of phase estimation and $R(\lambda^{-1})$ rotation. After the four photons pass through PBS$_3$, PBS$_4$ and PBS$_5$, we observed the Hong-Ou-Mandel type interference among the four photons \cite{HOM,Pan-RMP}. We measured the fidelity -- defined as the overlap of the experimentally produced state with the ideal one -- of the generated four-photon GHZ state \cite{otfried-PRA}. The measurements (see Fig.~S1) yield a state fidelity of 0.65(1), which exceeds the threshold of 50$\%$~\cite{Fidelity} by 15 standard deviations. This confirms the presence of genuine entanglement \cite{genuine} created during the quantum computation.

We have implemented the algorithm for various input vectors $|b\rangle$ which are varied by tuning the HWP in front of PBS$_3$.
In accordance with Fig.~\ref{fig1}b, the two registers should be projected to the state $|0\rangle$, the ancilla qubit to state $|1\rangle$. The output $|x\rangle$ is measured in some desired observable. In the experiment, each run of the algorithm is finished by a fourfold coincidence measurement where all four detectors fires simultaneously.

We characterize the output by measuring the expectation values of the Pauli observables $Z$, $X$, and $Y$ for each input state $|b\rangle$. Fig.~\ref{fig3} shows both the ideal (gray bar) and experimentally obtained (red bar) expectation values for each observable. To quantify the algorithmic performance, we compute the output state fidelity $F=\langle{x}|\rho_{x}|x\rangle$, where $|x\rangle$ is the ideal state and $\rho_{x}$ is the experimentally reconstructed density matrix of the output state from the expectation values of the Pauli matrices (see Fig.~S2). Compared with ideal outcomes, the output states have fidelities of $0.993(3)$ for $|b_{1}\rangle$, $0.825(13)$ for $|b_{2}\rangle$, and $0.836(16)$ for $|b_{3}\rangle$, respectively.

The difference in the performance for the three inputs is linked to the specific optical setup used in the experiment. The fidelity imperfections for $|b_{2}\rangle$ and $|b_{3}\rangle$ are caused by high-order photon emission events and post-selection in CNOT gates. However, in the case for $|b_{1}\rangle$, high-order photon emissions and post-selection do not give a negative contribution, giving rise to a near-ideal algorithm performance.


In summary, we have presented a proof-of-principle demonstration of the quantum algorithm for solving systems of linear equations
in a small-scale quantum computer involving four qubits and four entangling gates. We have implemented every subroutine at the heart
of the algorithm and characterized the circuit and algorithmic performances by the quantum state fidelities. The technique of coherently
controlling multiple qubits and executing complex, multiple-gate quantum circuits presents an advance on linear optics quantum
computation \cite{Kok2007,Pan-RMP} and allows to test other similar quantum algorithms such as solving
differential equations~\cite{Leyton2009,Berry2010} and data fitting~\cite{Wiebe2012}.

In principle, efficient quantum computation can be achieved using single-photon sources, linear optics, and single-photon detectors~\cite{KLM, Kok2007, 66efficiency}.
The current experiment, however, is still limited by a probabilistic single-photon source and inefficient detectors.
It can be expected that with ongoing progress on deterministic single-photon sources~\cite{yuming}, high-efficiency ($>$$93\%$) single-photon
detectors \cite{nistDetector}, and on-chip integration \cite{shorPhoton}, a larger-scale quantum circuit for solving more complex linear equations
can be implemented in the future.

\noindent \textit{note}: During the stage of manuscript preparation, we became aware of a related work~\cite{walther}.

\noindent \textit{Acknowledgement}: We thank Xi-Lin Wang, and Daniel James for helpful discussions. This work was supported by the National Natural Science Foundation of China, the Chinese Academy of Sciences and the National Fundamental Research Program (under Grant No: 2011CB921300). C.-Y.L. acknowledges Churchill College Cambridge and the Youth Qianren Program. N.-L.L acknowledges Anhui Natural Science Foundation. C.W. is supported by NSERC. M.G. is supported by the National Research Foundation and Ministry of Education in Singapore.

\end{document}